\documentclass[twocolumn,amsmath,amssymb,prb]{revtex4}
\usepackage{graphicx}
\usepackage{amsmath, amsthm, amssymb}
\usepackage{latexsym}
\usepackage{amssymb}
\usepackage{bm}
\usepackage{dcolumn}
\usepackage{epstopdf}
\usepackage{setspace}

\begin{document}
\title{Charge Induced Vortex Lattice Instability}

\author{A.M. Mounce,  S. Oh, S. Mukhopadhyay, and W.P. Halperin*}
\affiliation{Department of Physics and Astronomy, \\Northwestern  University, Evanston, IL 60208,
USA}
\author{A.P. Reyes,  and P.L. Kuhns}
\affiliation{National High Magnetic Field Laboratory, Tallahassee, FL 32310,
USA}
\author{K. Fujita, M. Ishikado, and S. Uchida}
\affiliation{Department of Physics, \\University of Tokyo, Tokyo 113-8656, Japan}
\date{Version \today}


\begin{abstract} {\bf It has been predicted that superconducting vortices should be electrically charged and that this effect is particularly enhanced for high temperature superconductors.\cite{kho95,bla96}  Hall effect\cite{hag91} and nuclear magnetic resonance (NMR) experiments\cite{kum01} suggest the existence of charge accumulation in the vortex core, but the effects are small and the interpretation controversial.   Here we show that the Abrikosov vortex lattice, characteristic of the mixed state of superconductors, will become unstable at sufficiently high magnetic field if there is charge trapped on the vortex core.  Our NMR measurements of the magnetic fields generated by vortices in Bi$_{2}$Sr$_{2}$CaCu$_{2}$O$_{8+y}$ single crystals\cite{che07} provide evidence for an electrostatically driven vortex lattice reconstruction with the magnitude of charge on each vortex pancake of $ \sim 2$x$ 10^{-3} e$, depending on doping, in line with theoretical estimates.\cite{kho95,kna05}} \end{abstract}

\maketitle

\newcommand{\dg}{^{\circ}}

\DeclareGraphicsRule{.tif}{png}{.png}{`convert #1 `basename #1 .tif`.png}

The behavior of quantized vortices in high temperature superconductors (HTS) affect many of their applications ranging from high current carrying wires\cite{she09} to electronic devices.\cite{mar09} Effective performance requires that the vortices be stable since their dynamics lead to dissipation originating from the electronic excitations within the vortex cores. Effort to understand the interactions between vortices has been focused on the circulating supercurrents that are characteristic of the vortex and generate local magnetic fields.\cite{cle91}  However, if the vortices are electrically charged there will also be an electrostatic interaction, possibly relevant to vortex stability.  We have performed model calculations showing that if vortices trap electrical charge, then the Abrikosov vortex lattice becomes unstable in sufficiently high magnetic fields where the magnetic interactions diminish and the electrostatics dominate.  Our NMR experiments confirm the existence of a magnetic field induced instability in the vortex structure of the highly anisotropic superconductor Bi$_{2}$Sr$_{2}$CaCu$_{2}$O$_{8+y}$ (Bi2212).

\begin{figure}[htp]
	\centering
		\includegraphics{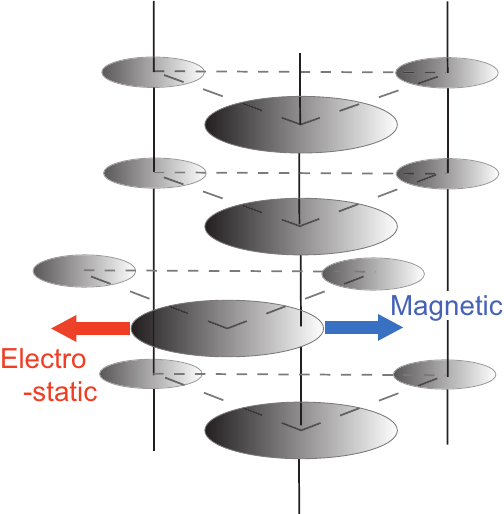}
		\caption{\textbf{Vortex pancakes with competing interactions.}  Attractive-magnetic and repulsive-electrostatic forces compete and change dominance as the magnetic field is increased.  The magnetic interactions between vortices in different planes is attractive preferring a line of vortex pancakes while the Coulomb interaction is repulsive.}
	\label{Fig1}
\end{figure}

A strong magnetic field normal to the conduction plane, $H$\textbar\textbar$c$, of Bi2212  penetrates in the form of quantized flux bundles, called pancake vortices.\cite{che07} They interact magnetically, with a repulsive interaction for pancakes within a plane, but attractive if they lie in different planes. The attractive force is responsible for aligning the vortex cores, one above the other, thereby forming flux lines as indicated in Fig.~1.  Perfect alignment is called the Abrikosov configuration.  However, the average attractive force decreases\cite{cle91} with increasing field as the vortex density increases and as the magnetic field modulation from nearby vortex supercurrents cancel each other more effectively.  In contrast, for charged vortex cores, the electrostatic force between pancakes on different planes increases at short range.   A complete picture of the balance of these forces requires that the nearby stacks of vortices be taken into account.  Clem calculated  the cost in magnetic energy for displacing an entire plane of vortices, relative to those in all other planes.\cite{cle91}  We follow this approach and compare the magnetic energy with the change in electrostatic energy when there is a charge on each vortex core.  Such displacements can become more favorable for charged vortices, decreasing the total energy as is illustrated in our calculations shown in Fig.~2.  At sufficiently high magnetic field the electrostatic and magnetic interactions balance and there is an instability.    The effect is evident in the inset of this figure at a field of 10 T  for a charge of $2.1 $x$ 10^{-3} e$, where the energy, as a function of vortex layer displacement in the $ab$ plane, becomes flat.  

  \begin{figure}[t]
	\centering
		\includegraphics[width=0.45\textwidth]{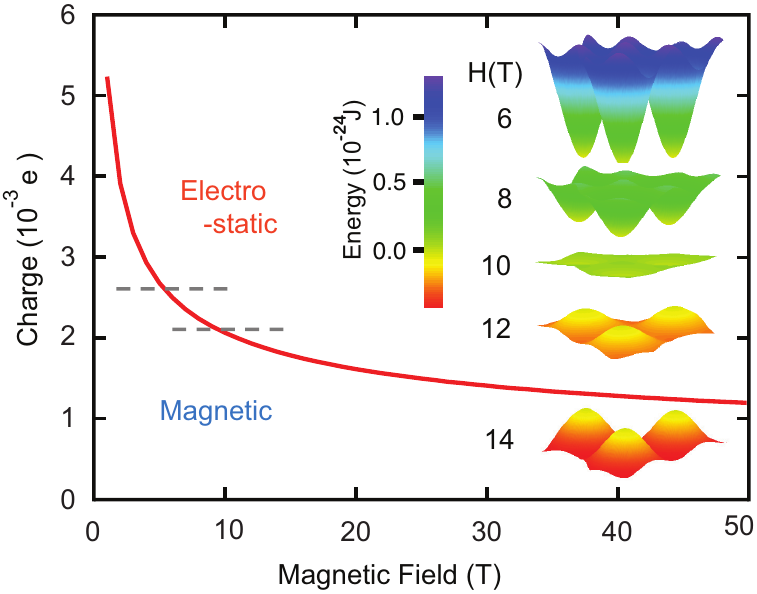}
		\caption{\textbf{Energy phase diagram.} Two separate regions in the charge-field plane correspond to dominance of charge or magnetic forces.  Below the red curve the magnetic energy is larger and above the red  curve the electrostatic energy dominates. The dashed lines are chosen to intersect this phase diagram at the magnetic field where the NMR second moments have a minimum, Fig.~3.  [Inset] At each magnetic field, for a charge of $2.1$x$ 10^{-3} e$, the change in total energy is shown for a rigid displacement of a layer of vortices in the $ab$ plane to various positions ranging over several vortex unit cells of the Abrikosov configuration. Softening of this energy landscape is evident at 10 T.}		
		\label{Fig2}
\end{figure}

  \begin{figure}[t]
	\centering
		\includegraphics[width=0.45\textwidth]{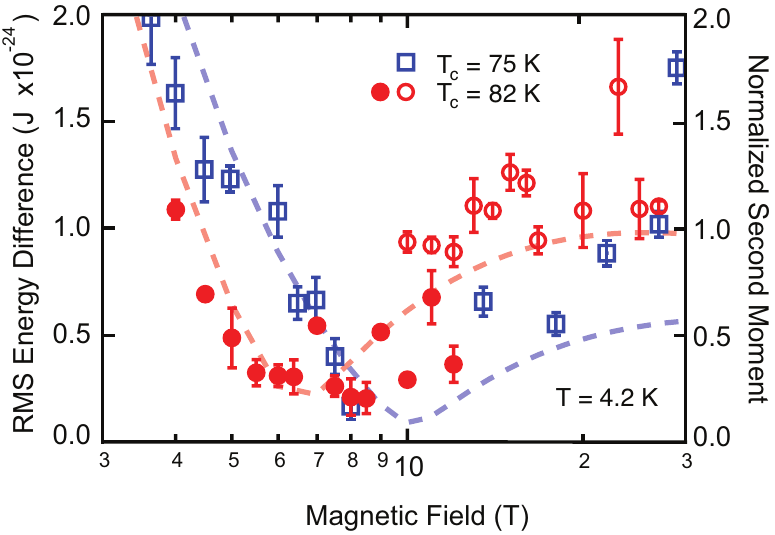}
		\caption{\textbf{Softness of the charged vortex structure.} The dashed curves are the root-mean-square energy cost for vortex displacements from the Abrikosov configuration perpendicular to the applied field for the effective vortex charges of $2.1$ and $2.5$ x $ 10^{-3} e$. These curves (left axis) are shown superposed on our NMR data, to demonstrate that the softening of the vortex-vortex interactions for these charges, occurs at the magnetic fields where the NMR second moment is a minimum. The second moments of the $^{17}$O NMR spectra (right axis) are shown  for three crystals independently processed by $^{17}$O exchange to give $T_{c} = 75 $ K (blue squares) and $T_{c} = 82 $ K (red circles, open and closed), at $T = 4.2$ K (error bars represent the background noise). The data are normalized to the theoretical value~\cite{bra91} for a Abrikosov vortex configuration at low magnetic field, $\sigma_0=0.00371\phi_0^2/\lambda^4$. An independently determined~\cite{che07}  Knight shift distribution that contributes 1.25 kHz/T  to the linewidth, has been subtracted.  At $H=8$ T this corresponds to 0.44  of the normalized second moment.  }
	\label{Fig3}
\end{figure}

The existence of charge trapped on the vortex core of a superconductor has been discussed extensively\cite{kho95,bla96,leb97,kol01,sim02,lip02,esc09} as a consequence of the Bernoulli potential for the electrostatic fields, and as early as 1961 by Fritz London.\cite{lon61}  Even for conventional superconductors the electrodynamics of vortices lead to a depletion of charge in the vortex core in order to balance  inertial and Lorentz forces.\cite{leb97} Additionally, if there is particle-hole asymmetry, it is favorable for charges in the core states to move to lower energy in the outer superconducting region in order to maintain constant chemical potential.\cite{kho95,bla96}  These different mechanisms produce a vortex charge per pancake\cite{kho95,esc09} of order $Q \sim e(\Delta/\epsilon_{F})^2$, a reduction from the electronic charge, $e$, by the ratio of the superconducting gap to the Fermi energy squared. This can be of order $10^{-6}e$ for low temperature superconductors but more favorably, $\sim 10^{-3}e$ for HTS, which have a larger gap.  Vortex charging has been studied theoretically for $d$-wave superconductors\cite{che02,kna05, zha08} coupled to antiferromagnetic (AF) order in the vortex core, and chiral triplet superconductors.\cite{esc09}  The experimental situation is not so well-established.  A sign change in the Hall effect\cite{hag91} has been interpreted\cite{kho95} as evidence for vortex charging.  However, sign changes have been recently ascribed to Fermi surface effects.\cite{leb07}  Comparison of the nuclear quadrupole resonance frequency and the satellite splittings in NMR has also been interpreted as evidence for vortex charging.\cite{kum01}  But the magnitude and sign of the charge, and the temperature dependence are in disagreement with BCS theory\cite{kum01, lip02} as well as $d$-wave theory.\cite{kna05}

\begin{figure}[t]
	\centering
		\includegraphics[width=.45\textwidth]{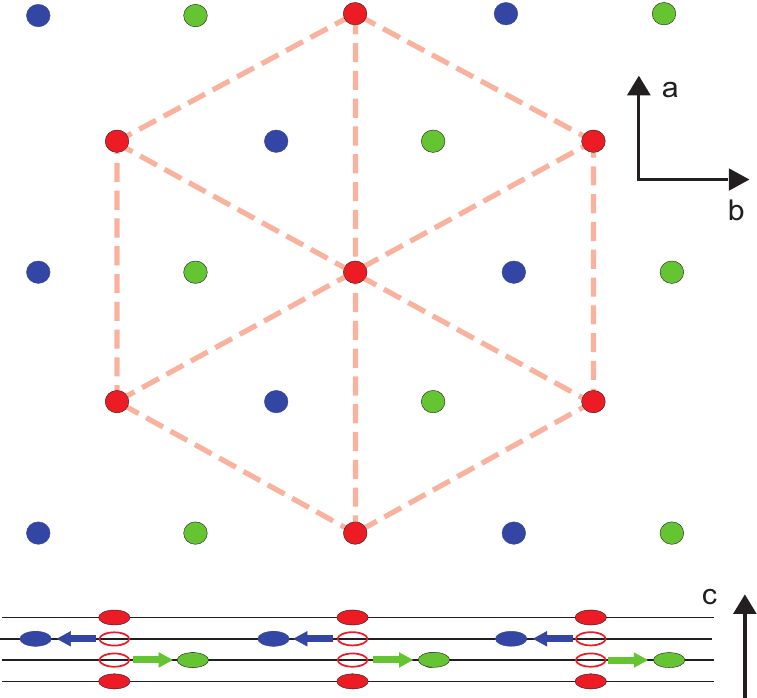}
		\caption{\textbf{Electrostatically driven phase change of the vortex lattice.}  The red symbols, both filled and open, represent vortex pancakes in the Abrikosov configuration.  At magnetic fields above the instability three inequivalent lattices appear (red, blue, and green).  The aspect ratio in the bottom portion of the figure matches the geometry of Bi$_{2}$Sr$_{2}$CaCu$_{2}$O$_{8+y}$ at $H=10$ T.  As the magnetic field increases the Coulomb potential becomes greater than the magnetic inter-layer coupling, making it favorable for pancake vortices to displace as far as possible from near neighbors in adjacent planes. }
		
	\label{fig4}
	
\end{figure}

We model the competing magnetic and electrostatic interactions by calculating the total energy per vortex for a rigid displacement of a single layer of pancake vortices in the $ab$ plane relative to the energy of the Abrikosov configuration.  Josephson coupling between planes is suppressed at high magnetic field by temperature fluctuations\cite{gla91} and can be ignored.  From Clem's calculation,\cite{cle91}  the magnetic coupling energy per vortex is given by, 

\begin{equation}
E_m({\bf{r}})=(H/\phi_0)\displaystyle\sum_{\bf{g}} U_0({\bf{g}})[e^{i{\bf{g}}\cdot {\bf{r}}}-1]
\end{equation}

where,
\begin{equation}
U_0(q)=-\phi_0^{2}s/4\pi \lambda^4(q^2+\lambda^{-2}).
\end{equation}

Here $H$ is the applied field, normal to the conducting planes; ${\bf{g}}$ is a reciprocal lattice vector of the Abrikosov vortex lattice in $q$-space; $\phi_{0}$ the flux quantum;  $\lambda = 220$ nm is  the in-plane penetration depth\cite{che07}; the bilayer spacing is $s$ = 15 $\AA$; and ${\bf{r}}$ the displacement of the plane of pancakes restricted to the vortex unit cell of the Abrikosov configuration (${\bf{r}}=0$).  The sums converge when taken to seven times the vortex spacing.  For the electrostatic energy we calculate the Coulomb potential of a periodic lattice and determine the energy required for the  displacement defined above.   The Coulomb potential for point charges is summed in direct space assuming a cut-off of $35$ times the inter-vortex spacing.  To maintain charge neutrality, there is a positive background charge spread throughout the unit cell, but when considering the energy difference of a displacement, its effect is canceled. The calculation of the energy landscape for the displacement of a single plane is displayed for various magnetic fields in the inset to Fig.~2.  At low magnetic fields the lowest energy corresponds to the Abrikosov configuration where all vortex cores line up, one above the other.  However, in high magnetic fields, the most favorable location is for vortices in the displaced plane to line up at the minimum magnetic field position equidistant from vortices in the plane above it (or below it), as sketched in Fig.~4. It is remarkable that at an intermediate field the energy landscape is essentially flat. This condition can be expressed quantitatively in terms of the root-mean-squared energy averaged over the unit cell, shown in Fig.~3.  The minimum in this $rms$-energy as a function of magnetic field defines the transition between magnetic and electrostatic order for a given vortex charge. The calculations are performed for different charges giving the phase diagram in Fig.~2. This calculated charge is really an {\it effective} charge which includes the Thomas-Fermi screening within a plane responsible for the suppression factor, $(\Delta/\epsilon_{F})^2$, as noted above.  But it does not allow for dielectric screening that will increase the charge by $\sqrt{\kappa}$ where the dielectric constant, $\kappa$, is a materials parameter $\sim  5$.\cite{lip02,mak07} Since Bi2212 is a bilayer system, the charge per conduction plane in a vortex is larger than the effective charge by $\sqrt{\kappa}/2\sim 1$.

\begin{figure}[t]
	\centering
		\includegraphics[width=0.45\textwidth]{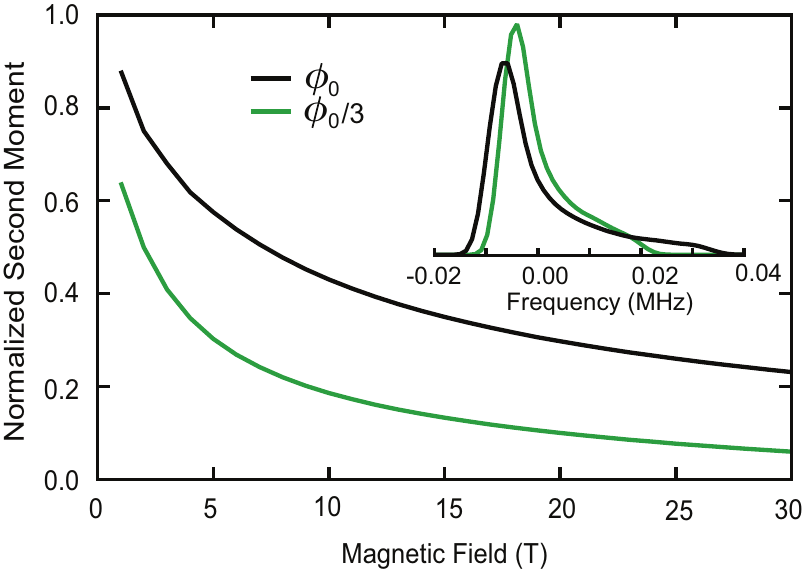}
		\caption{\textbf{Magnetic field dependence of the second moment for the vortex local-field distribution.} The black curve is calculated from Ginzburg Landau theory for the Abrikosov configuration.~\cite{bra97} The green curve corresponds to a tripled vortex unit cell which in our model is an electrostatically driven reconstruction of the vortex lattice obtained by changing $\phi_0$ to $\phi_{0}/3$ as suggested by Brandt~\cite{bra91} and depicted in Fig.~4.  There is a factor of 2 to 3 between the second moments of the two configurations at any given field. Such a vortex lattice reconstruction leads to a decrease in the second moment similar to our data, Fig.~3.[Inset] The calculated NMR spectra at $10$ T for both configurations of vortex pancakes are plotted with respect to their first moments.}	
		\label{fig5}
\end{figure}

The experimental NMR spectrum directly reveals the distribution of magnetic fields generated by vortices and their spatial correlations.  In the superconducting state the spectrum is simply proportional to the probability distribution of the local fields.  It is common\cite{mit01,che07} to compare experimental spectra with that expected for a regular array of vortex lines.  In this case the field distribution has a characteristic asymmetry and singularities associated with the minimum local field, the saddle point in the field distribution, and a maximum field at the vortex core, inset to Fig.~5.  The second moment of the spectrum, closely related to the square of the linewidth,  is a direct measure of the inhomogeneity of the fields from vortex supercurrents and can be interpreted in terms of the superconducting penetration length, $\lambda$, and the geometry of the vortex array.  If the vortex pancakes are disordered within the conduction plane, the distribution is broadened.  In contrast, disorder or reconfiguration of vortices from plane to plane decreases the linewidth.\cite{bra91}  From our NMR experiments we find an unusually abrupt decrease in the linewidth with increasing applied magnetic field for $^{17}$O NMR measured at $T=4.2$ K at the O(1) oxygen site in the copper-oxygen conducting plane of Bi$_{2}$Sr$_{2}$CaCu$_{2}$O$_{8+y}$.   The magnetic field dependence of the second moment of the spectrum is shown in Fig.~3. The magnitude of the second moment at lower fields, $H\lesssim5$T, is of order that expected from an Abrikosov configuration. We interpret the linewidth collapse with increasing field  as a transformation to a new, field-induced, vortex state.

The abrupt decrease of the NMR linewidth with increasing magnetic field for several over-doped crystals ($T_{c} = 75 $ K, $82$ K), Fig.~3, provides consistent experimental evidence that charged vortices can be a mechanism for vortex structure instability. Additionally, our results indicate that this instability moves to higher magnetic fields with increased doping, corresponding to a smaller vortex charge as suggested by calculations for a $d$-wave superconductor.\cite{kna05} In related work, Gurevich found that if a single vortex line is charged, it becomes unstable to chiral distortion.\cite{gur10}

In our model the Abrikosov configuration is transformed to an $A$-$B$-$C$-type of tripling of the vortex unit cell along the $c$-axis as sketched in Fig.~4. At high magnetic fields a pancake vortex seeks out a position that is anti-correlated with vortices in planes above and below it.  From the minimum NMR linewidth in Fig.~3 we find the magnetic field for this transition. Then according to the phase diagram in Fig.~2, the effective charges per pancake are $2.1$ and $2.5$ x $ 10^{-3} e$ for the crystals $T_{c} = 75 $ K and $T_{c} = 82 $ K, respectively.  This new ordering maintains the intra-plane vortex-vortex correlations but the spacing of vortex cores on adjacent planes is decreased by $\sim 1/\sqrt{3}$.  This leads to a decrease in the distribution of local magnetic fields as pointed out by Brandt\cite{bra91} and a corresponding decrease in the second moment of the NMR spectrum as we show in Fig.~5.  Raw spectra and further experimental details are presented in the Supplementary Materials section.  On close examination the NMR spectra at low field have two distinct components which merge with increasing field. The composite spectra have an overall shape rather similar to the superposition of the calculated spectra shown in the inset to Fig.~5.  We infer that it is likely that the vortex rearrangement proceeds continuously over a range of magnetic field and completes at the field for which the spectrum width is a minimum. The minimum is a balance between the abruptly decreasing vortex contribution to the local field distribution and an increasing paramagnetic contribution with increasing applied field that we have followed up to as high as 30 T, Fig.~3.  We identify this high field behavior with the magnetization induced by bound states in the vortex core.\cite{mit01}\\

\section{METHODS}

	Our samples were prepared by exchanging in 1 bar of flowing $^{17}$O (70-90\% enriched) at $600^{\circ}$ C for 48 h  followed by annealing for 150 h at $450^{\circ}$ C.  Our $82$ K data in Fig.~3 are from two separate samples from the University of Tokyo.  The closed circles correspond to three single crystals of Bi-$2212$, stacked along the c-axis, with a total mass of $20$ mg.  The sample represented by the open circles has a similar total mass, though the crystals were cleaved to improve signal-to-noise ratio. The squares are data from  a previous experiment\cite{che07} with crystals, $T_c= 75$ K, provided by Argonne National Laboratory with similar processing conditions.
	
	NMR measurements were performed in magnetic fields from $4$-$30$ T at a temperature of $4.2$ K at Northwestern University and the National High Magnetic Field Laboratory in Tallahassee, Florida.  We have shown that at this temperature the vortex structure is frozen at all magnetic fields.\cite{che07} Spectra were taken using a frequency sweep technique with a typical $\pi$/2 pulse of 4 $\mu$s.  Before taking data at any particular magnetic field, the sample was heated above $T_c$ for $10$ min and then quench-cooled to $4.2$ K. This method produces narrow spectra, with minimal flux pinning, and was found to give consistent results through multiple field cycles.	 In general flux pinning is weaker at higher magnetic fields, and if it exists can be identified using field-sweep NMR.

\section{ACKNOWLEDGEMENTS}	
	We thank E.H. Brandt, L.N. Bulaevskii, C.A. Collett, M. Eschrig,  M.R. Eskildsen, W.J. Gannon, A. Gurevich, A.E. Koshelev, Jia Li, V.F. Mitrovi\'c, J. Pollanen, and J.A. Sauls for helpful discussions.  This work was supported by the Department of Energy, contract DE-FG02-05ER46248 and the National High Magnetic Field Laboratory, the National Science Foundation, and the State of Florida.


\section{Author Contributions}

Experiments were carried out by A.M.M., S.O., S.M., W.P.H., A.P.R. and P.L.K.  Samples were provided by K.F., M.I., and S.U.

\section{Author Information}

Correspondence and requests for materials should be addressed to w-halperin@northwestern.edu.  

\section{Competing financial interests}

The authors declare no competing financial interests

\section{Supplementary Materials}
\subsection{Bi-2212 $^{17}$O NMR}

\begin{figure}[htp]
	\centering
		\includegraphics[width=.45\textwidth]{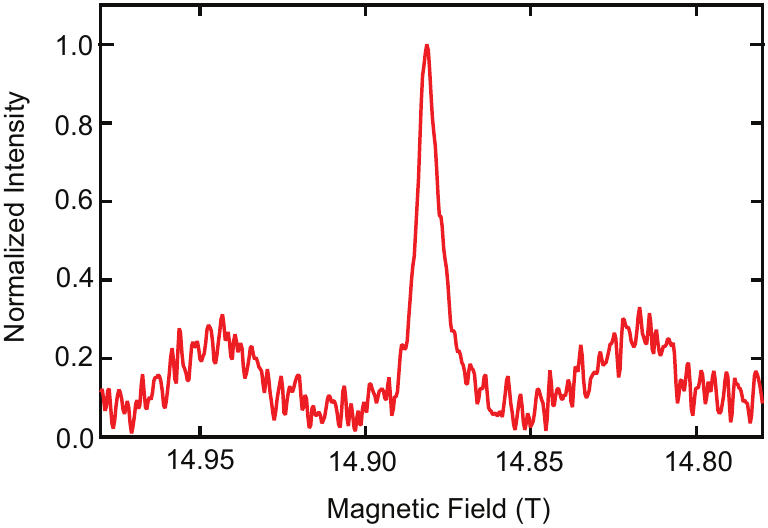}
		\caption {\textbf{The $^{17}$O NMR spectrum of the T$_c$ = 82 K sample O(1) site at a frequency of 86 MHz and temperature 4.2 K.} The central transition $\langle \pm 1/2 \leftrightarrow \mp 1/2 \rangle$ is shown along with the first pair of quadrupolar satellites, $\langle \pm 3/2 \leftrightarrow \pm 1/2 \rangle$, with a quadrupolar splitting of $\Delta\nu_Q= 63$ kHz. For the orientation of our sample, H $\parallel c$-axis, the central transition has no quadrupolar broadening. }
\end{figure}

With multiple stoichiometric locations and quadrupolar resonances, the full oxygen NMR sectrum of Bi-2212 is complex.  The main peak in Fig.~6 is the central transition, $\langle \pm 1/2 \leftrightarrow \mp 1/2 \rangle$, of the $^{17}$O spectrum of Bi-2212 from the O(1) site. This site  is understood to be the oxygen resonance from the superconducting CuO$_2$ planes.  The O(2) resonance from the oxygen in the insulating SrO planes is completely saturated in our measurements at 4.2 K and is not discernible in Fig.~6. The oxygen in the BiO plane is unobservable due to structural disorder in this plane. The O(1) spectrum, in addition to the central transition, is split into two different pairs of quadrupolar satellites for the $\langle \pm 3/2 \leftrightarrow \pm 1/2 \rangle$ and $\langle \pm 5/2 \leftrightarrow \pm 3/2 \rangle$ transitions, the latter are just outside of the frequency range  presented in Fig.~6, see Ref 5 for the full spectrum.

\begin{figure}[htp]
	\centering
		\includegraphics[width=.45\textwidth]{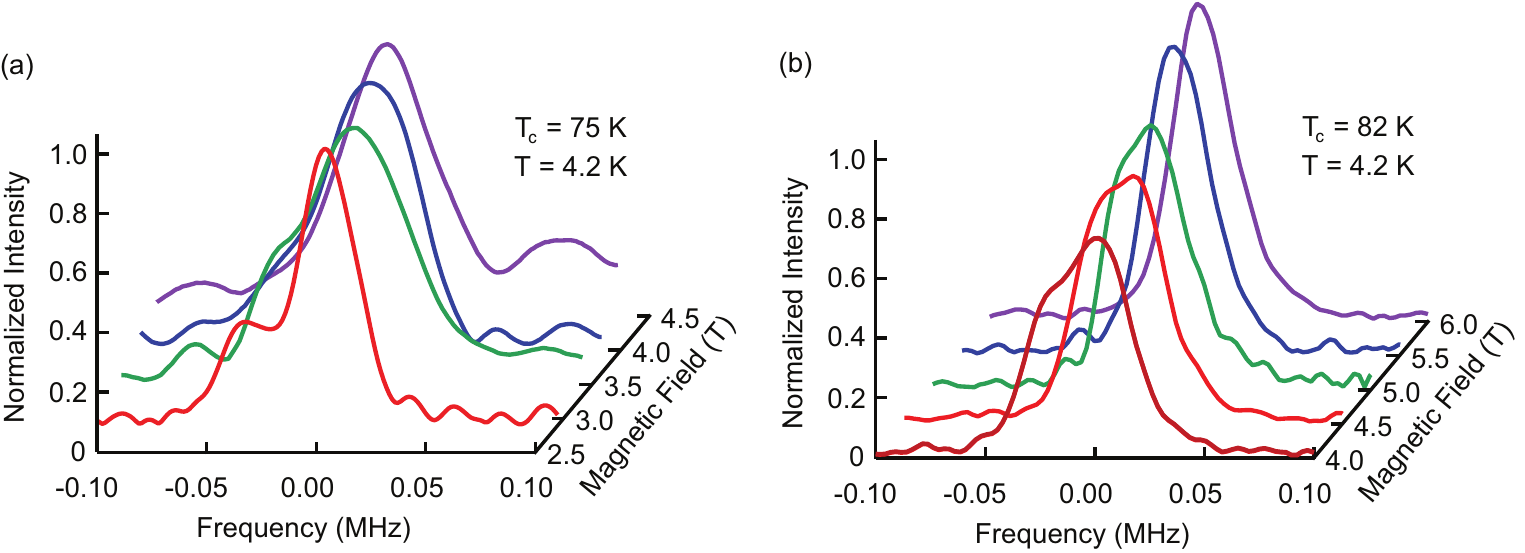}
		\caption{\textbf{Field evolution of the central O(1) NMR spectra for crystals with T$_{c}$ = 75 and 82 K.}  Each spectrum was recorded after the sample was field cooled through the transition temperature to avoid flux pinning as described in the Methods Section.  The lineshape of the spectra at these low fields appears to have two components that merge near $H=4.5$ and 6 T, respectively, for these two crystals.}

\end{figure}

As the sample is cooled from the normal state, through the superconducting transition, and finally to the vortex solid phase, the NMR spectrum shows a clear temperature dependence described in Ref. 5.    At relatively high magnetic fields and at temperatures well below T$_c$, vortices form a solid phase; the spectrum broadens and becomes asymmetric as would be expected for the Abrikosov vortex lattice, Ref. 5. 

\subsection{Multicomponent Spectra}

The broad spectra at low fields, for which the second moments are shown in Fig.~3, appear to be composite spectra with two principal components.  These spectra  are displayed in Fig.~7 for the two crystals with, (a)  $T_{c}$ = 75 and, (b) 82 K.  The spectra narrow as the lower frequency contribution decreases in intensity with progressively increasing magnetic field.  The sample with $T_{c} = 75 $ K  has  two components which are somewhat narrower and spread further apart, as compared with the crystal with $T_{c} = 82 $ K.   We note that the calculated field distribution for two vortex states: the Abrikosov configuration of aligned vortices and that for the anti-aligned vortex pancakes, have a shift in frequency in addition to a different magnetic field distribution, as displayed in the  inset to Fig.~5. Their superposition is similar to the measured spectra in Fig.~7, suggesting that the instability observed  is a crossover from an inhomogeneous distribution of such vortex states to a more stable distribution at high magnetic fields. This complex behavior is not captured by our model for competition between electrostatic and magnetic interactions between vortex pancakes in different conduction planes.  Nonetheless, it is qualitatively consistent with the general picture which we propose of a field-induced instability attributable to vortex charge.

\end{document}